**The Effect of COVID Restriction Levels on Shared Micromobility Travel Patterns: A Comparison between Dockless Bike Sharing and E-Scooter Services**

**Marco Diana**
Department of Environment, Land and Infrastructure Engineering
Politecnico di Torino, 24 Corso Duca degli Abruzzi, 10129 Turin, Italy
Email: marco.diana@polito.it
ORCiD: 0000-0003-3832-6834

**Andrea Chicco**
Department of Environment, Land and Infrastructure Engineering
Politecnico di Torino, 24 Corso Duca degli Abruzzi, 10129 Turin, Italy
Email: andrea.chicco@polito.it
ORCiD: 0000-0003-0266-4735

Word Count: 6825 words + 2 tables = 7325 words

*Paper presented at the 101st Annual Meeting of the Transportation Research Board, Washington, D.C., January 9-13, 2022*

*Submitted July 30, 2021*

**ABSTRACT**

The spread of the coronavirus pandemic had negative repercussions on the majority of transport systems in virtually all countries. After the lockdown period, travel restriction policies are now frequently adapted almost real-time according to observed trends in the spread of the disease, resulting in a rapidly changing transport market situation. Shared micromobility operators, whose revenues entirely come from their customers, need to understand how the demand is affected to adapt their operations. Within this framework, the present paper investigates how different COVID-19 restriction levels have affected the usage patterns of shared micromobility.

Usage data of two dockless micromobility services (bike and e-scooters) operating in Turin (Italy) are analyzed between October 2020 and March 2021, a period characterized by different travel restriction levels. The average number of daily trips, trip distances and trip duration are retrieved for both services, and then compared to identify significant differences in trends as restriction levels change. Additionally, related impacts on the spatial dimension of the services are studied through hotspot maps.

Results show that both services decreased during restrictions, however e-scooters experienced a larger variability in their demand and they had a quicker recovery when travel restrictions were loosened. Shared bikes, in general, suffered less from travel restriction levels, suggesting their larger usage for work and study-related trip purposes, which is confirmed also by the analysis of hotspots. E-scooters are both substituting and complementing public transport according to restriction levels, while usage patterns of shared bikes are more independent.

## INTRODUCTION

The spread of the coronavirus pandemic has strongly affected the majority of transport systems in virtually all countries. Containment policies had indeed a significant repercussions on transportation, with a stronger impact on the demand and supply of services for passenger mobility (*1*). In particular, public transport systems have been the most negatively affected (*2–5*), with a remarkable reduction in their usage, which ranged between 70% and 90% in European countries (*3*). Lockdown restrictions have also negatively impacted the usage of shared mobility services, which resulted in suspending or even closing operations. In United States, shared micromobility services, namely docked and dockless bikes, e-bikes and e-scooters, which had exponentially grown since their first appearing in 2017 (*6*), declined from 2019 to 2020 because of the spread of the pandemic (*7*). In particular, dockless bike sharing declined from 71 systems in 2019 to 49 systems in August 2020, while e-scooter systems declined from 239 systems in 2019 to 173 systems in August 2020 (*7*).

As a consequence, the financial equilibrium of many public transport and shared mobility services has been jeopardized, putting many of them at risk, while an increase the use of individual transport means (*8, 9*), either motorized (car, motorcycle, and moped), human-powered or electrically-assisted (walk, bike, e-bike, and e-scooter) has been recorded (*10, 11*). The decrease in patronage is particularly challenging those services whose revenues entirely come from the farebox and that are not yet well established, since they started their operations only recently and the service concept itself is not familiar to a sizable proportion of citizens. Shared micromobility services seem therefore particularly hit by the consequences of the pandemic.

The term "shared micromobility" refers to those services based on the shared use of micro vehicles, that in turn are defined by the International Transport Forum as: "[..] *vehicles with a mass of no more than 350kg (771 lb) and a design speed no higher than 45 km/h.*" (*12*), that can be either human-powered or electrically assisted. Bikes and e-bikes (docked or dockless), mopeds, and standing e-scooters are typical examples of micro vehicles belonging to the shared micromobility world.

To face the pandemic, shared micromobility operators need to understand how to adapt their services when travel restrictions are put into place. Model-based forecasting techniques that are commonly used to build transport planning scenarios are difficult to use to this effect, since standard simulation tools calibrated and validated in "ordinary" conditions are likely to produce biased results in such an unprecedented situation. On the other hand, restriction policies are frequently adapted almost real-time according to observed trends in the spread of the disease, resulting in a rapidly changing transport market situation. Furthermore, the effect of such policies might somewhat overlap with that of weekly or seasonal variations of the demand (*13*), and might also change according to the characteristics of different services. Clearly, all these factors make it difficult to anticipate how service operations are affected through the different phases of the pandemic.

Understanding how shared micromobility services operations are affected by different levels of restrictions and according to the specific service is therefore a tall task. The present paper aims at giving a contribution by investigating how different COVID-19 restriction levels that were implemented in the city of Turin (Italy) between October 2020 and March 2021 have affected the usage patterns of two micromobility services, namely dockless shared bikes and shared e-scooters, in a different way. A summary of previous studies on shared micromobility services and how COVID impacted their operations is presented in the following section, whereas the experimental context and the data analysis methods are later presented. The paper then presents the results and summarizes the main findings in the conclusions.

## SHARED MICROMOBILITY BACKGROUND

There is an increasing body of literature about shared micromobility, in particular dockless bike sharing and e-scooters, that ranges from safety issues (*14*) to operations and business strategies (*15, 16*), environmental impacts (*17*), effects on social inclusion (*18, 19*), mode choice (*20*) and the wider impacts on other transport alternatives (*21–23*). Many researches are focusing on understanding the demand for such transport means. Besides understanding who is using shared micromobility and why, which is generally addressed through mobility surveys (*22–24*), other studies aim to clarify where and when such services are used (*25–27*). To address such research questions, several studies have exploited the operational data directly provided by service operators (*26*), made available by city administrations (*27, 28*), or scraped from the openly available advanced programming interface (API) of such services (*25*). This study falls in the latter category and therefore it analyses only trip origin and destination locations (*25, 28*), while complete GPS tracks that may be used to understand trip patterns (*29*) are missing.

Many studies based on similar operational data consider one service typology at once (*28, 30, 31*), compare docked and dockless bike sharing systems (*32*), sometimes jointly analyze docked bike sharing and dockless e-scooters (*25*), while only a handy of studies compare more than two service typologies (*20*). Therefore, little is known about the similarities or (differences) in usage patterns of different micromobility services. Additionally, most of these studies is carried out in United States (*25, 27, 28*) whereas other geographical areas are largely missing.

In Italy, dockless bike sharing services firstly appear in late 2017 (*33*) whereas first shared e-scooters became available in late 2019, through pilot programs in some cities, just before the lockdown period (*9*). Despite the number of shared e-scooter has grown almost 600% from December 2019 to September 2020 in the whole country (*9*), no research focusing on their usage within Italian cities have been published yet.

Lastly, very limited research has been carried out to understand how usage patterns of different micromobility services have changed in response to different restriction levels dictated by the COVID-19 pandemics. To date, to the best Authors' knowledge, only a case study conducted in Zurich, Switzerland analyzed the changes in micromobility usage in response to the COVID-19 pandemic, by comparing three micromobility services (docked bikes, docked e-bikes, and dockless e-bikes) before and during the lockdown periods (*34*). Comparing the two periods, it was found that the number of trips were significantly reduced due to pandemic. In particular, trip volumes decreased remarkably during peak hours on workdays, while less significant changes were observed on weekends compared to the period before the lockdown; work from home was indicated by the authors as the most influential driving factor for the temporal change. Nevertheless, during the lockdown period the proportion of trips related to home, park, and grocery activities increased, while the proportion of shopping and leisure activities decreased (*34*).

Differently from this case study, which compared the usage of micromobility services before and during the lockdown periods, our focus is on the changes in usage patterns that occurred under different levels of restrictions in the post-lockdown phase. As discussed in the introduction, we believe that this different perspective would be useful given the fact that both travel restrictions and the consequences of the pandemics on the travelers' mindset are not likely to completely disappear in the near future. Understanding which shared micromobility services are most affected by specific policy measures is therefore important to ensure that such services can continue their operations in these challenging times.

## EXPERIMENTAL CONTEXT

To evaluate the effects of COVID-19 restrictions on shared micromobility travel patterns, we exploited the data collected from the available APIs of two micromobility services (one providing a dockless bike and the other a dockless e-scooter sharing service) operating in the city of Turin, Italy. Data collection activities covered a large time span (147 days) starting from October 19, 2020 and ending on March 14, 2021.

### Travel restriction levels

Within the analyzed period, several important limitations to personal mobility dictated by the COVID-19 health emergency have been introduced. In particular, to contrast the appearance of a COVID-19 second wave (*35*), Italian Regions were classified into one of the four groups, namely "Red", "Orange", "Yellow", and "White". Each group was associated to different risk scenarios to which corresponded specific restrictive measures. Since the entry into force of this classification system (which occurred on November 4, 2020[1]) risk scenario attribution of all Regions was periodically updated on the basis of ordinances issued by the Italian Ministry of Health.

The main restrictions related to each level are summarized in **TABLE 1**, according to the definitions that were valid until the end of March 2021. The last column of the table shows the number of days in which the study area was classified into one of the four groups within the observation period, or similar restrictions were put in place until the 3rd of November 2020. The sum of the number of days displayed in the last column is less than the above observation period, since nine days in which the collected data were incomplete had been discarded.

**TABLE 1 Travel Restriction Levels Definition**

| Restriction Level | Travel Restriction | Days within data collection period |
|---|---|---|
| White (W) | • All commercial and educational activities are open with specific limitations on closed spaces (wearing masks and reduced capacity).<br>• Curfew: Not applicable. | 7* |
| Yellow (Y) | • Education: kindergartens, primary schools, and secondary schools are open.<br>• Shops: open.<br>• Restaurants and bars: sit-down services open till 6 PM (delivery till 10 PM). Entertainment (gyms, cinemas, museums, etc.): closed.<br>• Visiting relatives/friends: allowed.<br>• Curfew: from 10 PM to 5 AM. | 54 |
| Orange (O) | • Education: only kindergartens and primary schools are open.<br>• Shops: open.<br>• Shopping malls: closed on weekends and public holidays except for grocery stores, pharmacies, other shops selling essential goods inside the shopping mall.<br>• Restaurants and bars: closed, only delivery till 10 PM.<br>• Entertainment (gyms, cinemas, museums, etc.): closed.<br>• Curfew: from 10 PM to 5 AM. | 43* |
| Red (R) | • Only motivated trips are allowed (work, health, groceries).<br>• Education: only kindergartens and primary schools are open.<br>• Shops: all closed except for grocery stores, pharmacies, newsstands, shops selling essential goods.<br>• Shopping malls: closed<br>• Restaurants and bars: closed, only delivery till 10 PM.<br>• Entertainment (gyms, cinemas, museums, etc.): closed.<br>• Visiting relatives/friends: not allowed.<br>• Curfew: all day. | 34 |

* Including some days before the official entry into force of the Decree in which the risk scenarios have been judgmentally assigned by the authors according to the enforced restrictions.

---

[1] http://www.salute.gov.it/portale/nuovocoronavirus/dettaglioContenutiNuovoCoronavirus.jsp?lingua=english&id=5367&area=nuovoCoronavirus&menu=vuoto - Accessed March 14th, 2021

**Shared micromobility data**

During the analyzed period, the APIs of the two services were enquired with a set of requests sent out every 60 seconds to get shared bikes and e-scooters availability over the whole operational area of Turin. Every valid request's response contained a list of records, each one representing a bike or an e-scooter not in use at the time of the request. Each record contained information about the unique ID of the vehicle, its localization (latitude and longitude coordinates), the timestamp of the request, and the level of charge of the battery (for e-scooters). The resulting dataset counts about 300M records. Being a collection of available vehicles, the dataset did not contain any user information or user's type of subscription (single trip or pass).

To identify potential trips, records were firstly sorted by ID and timestamp and then consecutive records with common ID and GPS positions were discarded, since they represent bikes or e-scooters parked in the same location for that period. Meanwhile, consecutive records with the same ID and GPS position even marginally different were retained and combined; this resulted in about 5.7M displacements. Then, the Euclidean distance between each pair of origin and destination points were computed, together with the battery level consumption (only for e-scooters). However, most of the displacements showed a distance shorter than 10m, which was more likely to be associated with GPS accuracy issues related to urban canyons and the unavailability of satellites (*29*, *30*) than real trips. Therefore, a data cleaning procedure has been applied to filter out unrealistic trips that meet the following criteria:

- Euclidean distance between trip ends <50 m or >15 km.
- Trip duration <2 minutes.
- Trip duration >60 minutes for e-scooters.
- Trip duration >120 minutes for dockless BS.
- Speed > 20km/h.
- Power consumption <0 (the battery has been recharged).

It is worth noting the use of a different upper bound for the trip duration, which was set according to the maximum number of minutes per ride included in pre-paid passes offered by each service, and that may influence the results, especially when comparing the two services.

Additionally, as previously mentioned, nine days with incomplete data due to technical issues (server breakdown or connection errors) were furtherly discarded to avoid including potentially biased information. As a result, 138 days under different COVID-19 restriction levels accounting for 79,096 trips are finally retained.

**METHODS**

In order to understand to which extent different restriction levels might have impacted the usage patterns of the two micromobility services, trips detected through the above described procedure are grouped according to the service typology (shared bikes or shared e-scooters) and the COVID-19 restriction level ("White", "Yellow", "Orange", and "Red"). A further a distinction is made between days of the week (weekday or weekend) since both dockless bike sharing and e-scooter sharing usage patterns generally differ between such days, as observed in previous studies (*25*, *36*).

For each group, the average number of daily trips, distances between origins and destinations and trip durations are calculated and tested for significance through the non-parametric Kruskal-Wallis test (*37*), since it makes relaxed assumptions regarding distributions compared to the parametric ANOVA. Indeed, the considered variables do not follow a normal distribution according to preliminary Shapiro-Wilk normality tests. If a significant statistical difference is detected, a Dunn post-hoc test is performed to understand among which groups the statistical difference exists.

Additionally, to explore the spatial dimension of such micromobility services under different restriction levels and identify differences in core operating areas, origin and destination locations of the grouped trips are analyzed through a density-based algorithm for hotspot identification, the Kernel Density Estimation (KDE) (*38*).

In particular, the two-dimensional KDE tool available through the software QGIS (*39*) is used to compute the density surface and to store its values in a raster dataset with a spatial resolution of 5 m by 5 m. A quartic shape kernel function and a bandwidth of 500m were used for this analysis. For more details about the choice of the bandwidth value, the reader is referred to Chicco and Diana (*40*).

Heat maps are obtained by coloring the computed density surface with different colors depending on density intervals, which are set according to the elevation of the density surface with respect to the mean value (M). Five density intervals are considered: below the mean (light blue), equal or slightly above the mean value (green) and above the mean value plus 1, 3, and 5 times the standard deviation (SD) (yellow, orange, and red, respectively). Heat maps of usage of each service under different COVID restriction levels are then visually compared, also considering land use patterns.

**RESULTS**

**Daily trips time series within the observation period**

Shared micromobility usage trends in the city of Turin during the period under analysis are shown in **Figure 1**. In particular, the picture focuses on the temporal trends in relation with the different restrictions, showing the weekly moving average (MA) of the number of daily trips for both services, along with the mean number of trips within each time period characterized by constant travel restrictions.

Regardless of the different average number of daily trips of each service in absolute terms, both services are characterized by a sharp reduction of the travel demand between October and mid-November 2020. This can be directly due to the increased level of restrictions, which progressively moved from white to red. On the contrary, the two services show opposite trends between November 15 and December 5, where shared e-scooters are increasingly being used and shared bikes maintain a steadily decreasing trend. The marked reduction shown in both trends during orange restrictions in December (from the 5[th] to the 10[th]) coincides with a week of bad weather conditions, which generally affects this kind of services (*41*). Apart from this fluctuation, considering the whole period from mid-November to the end of December, e-scooter trips show an increasing trend, while bike sharing trips maintain a steady decreasing trend. Interestingly, the period before Christmas was characterized by an easing of the restrictions (yellow level), which allowed many shops to open for the traditional Christmas shopping. The peak shown by e-scooters might be connected to the fact e-scooters are more used to satisfy maintenance and discretionary trips such as shopping, compared to dockless bike sharing. Therefore, they could be more sensitive to restrictions in commercial activities.

During the last week of December and the first days of January, more severe restrictions were again applied, which probably contributed to the drop of e-scooter and, although to a lesser extent, bike sharing usage. It is worth mentioning that such period is generally characterized by an atypical travel demand; it is however difficult to understand the causal relationship between such events and the observed usage trends.

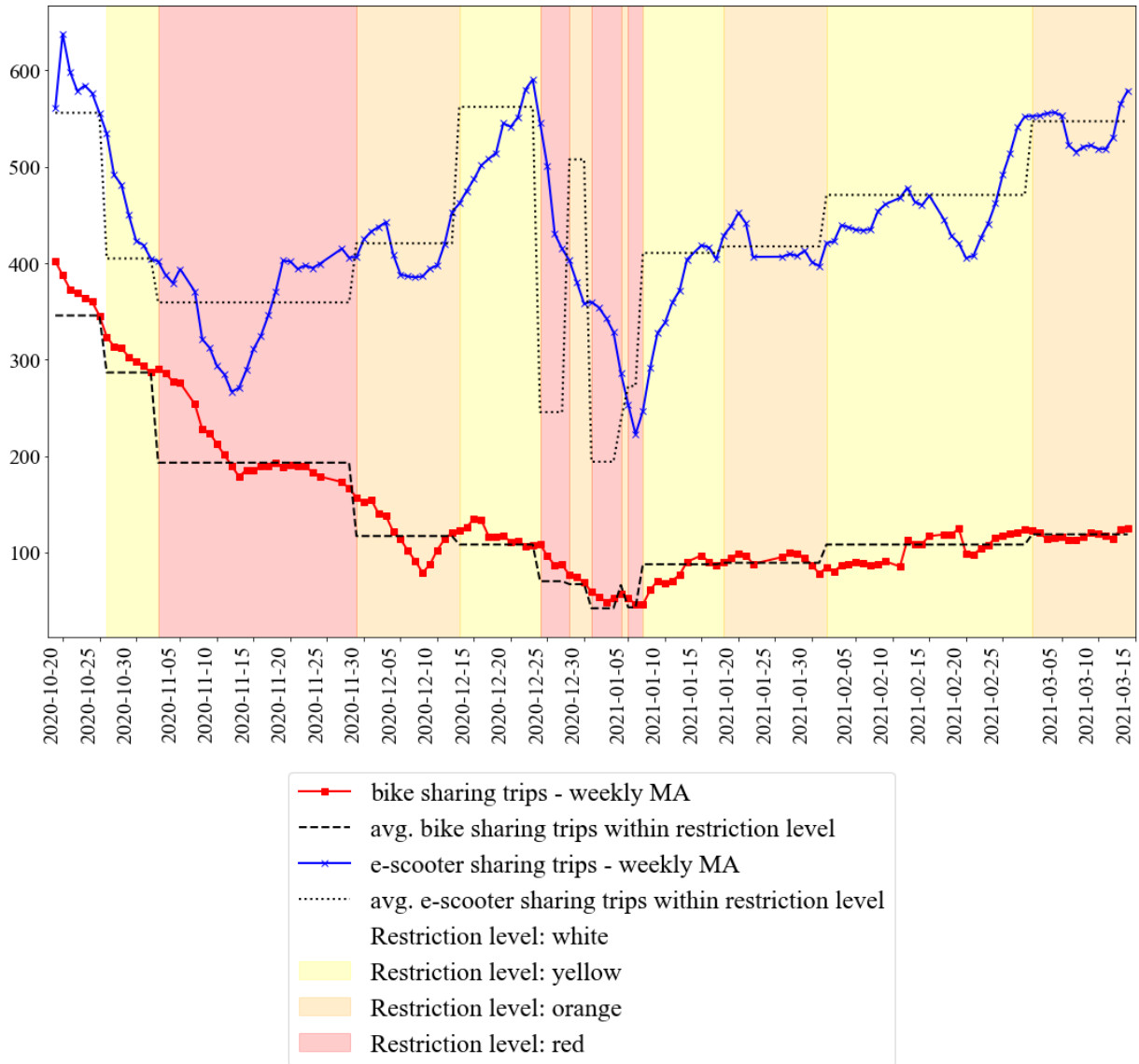

**Figure 1 Time series of the daily number of trips by bike sharing and by e-scooters within the considered period**

Lastly, both micromobility services show increasing usage trends from mid-January onwards; yet e-scooters usage increases at a faster rate compared to bike sharing. It is remarkable that the average number of trips within each period characterized by the same level of restrictions (dotted and dashed lines of **Figure 1**) monotonically increases in 2021, irrespective of the restriction level, which seems to indicate that weather conditions had a more substantial effect compared to different restriction levels.

E-scooters demand appears more sensitive to both COVID-related travel restrictions (fluctuations between neighboring periods) and other exogenous factors (fluctuations within each period) compared to dockless shared bikes. Bike sharing usage, on the other hand, shows a smoother trend, relatively less affected by COVID-19 restrictions which may underlie a different usage. Indeed, travel restrictions as described in **TABLE 1** had a more severe impact on non-commuting rather than on commuting and study-related trips.

**Changes in the number of trips, travel distance and duration according to different restriction levels**

Descriptive statistics about the average number of daily trips, the average Euclidean distance between origins and destinations (that is a reasonable lower bound of the trip distance that is indicated in the table) and average trip duration for dockless bike sharing and e-scooter services are presented in **TABLE 2**. The figures reported in each row refer to a different COVID-19 restriction level, as defined in **TABLE 1**, while the last row represents average values over the entire period. To better highlight the changes occurring under different level of restriction, the percent variation of each quantity with reference to the white restriction level are reported in parentheses. Statistically significant differences according to Kruskal-Wallis tests are identified by superscripts pointing to the different periods.

**TABLE 2 Descriptive Statistics of Dockless Bike Sharing and E-Scooter Use for Different Travel Restrictions**

| | Dockless bike sharing | | | | | | e-scooters | | | | | |
|---|---|---|---|---|---|---|---|---|---|---|---|---|
| | Daily trips | | Trip distance (km) | | Trip duration (minutes) | | Daily trips | | Trip distance (km) | | Trip duration (minutes) | |
| | wd | we | wd | we | wd | we | wd | we | wd | we | wd | we |
| W | 364[Y,O,R] | 300[Y,O,R] | 1.43[Y,R] | 1.25[Y,O,R] | 22 | 23[Y,O] | 584[R] | 485[R] | 1.54[Y,O,R] | 1.65[Y,O,R] | 17[Y,O,R] | 18[Y,O,R] |
| Y | 143[W] (-61%) | 94[W] (-69%) | 1.35[W,R] (-6%) | 1.35[W,R] (9%) | 22 (0%) | 26[W,R] (13%) | 487[R] (-17%) | 433[R] (-11%) | 1.47[W,O] (-5%) | 1.39[W] (-16%) | 15[W] (-12%) | 16[W] (-11%) |
| O | 110[W,R] (-70%) | 92[W] (-69%) | 1.39[R] (-3%) | 1.38[W,R] (11%) | 23 (5%) | 26[W,R] (13%) | 471[R] (-19%) | 434[R] (-11%) | 1.43[W,O] (-7%) | 1.43[W] (-13%) | 15[W] (-12%) | 16[W] (-11%) |
| R | 171[W,O] (-53%) | 112[W] (-63%) | 1.30[W,Y,O] (-9%) | 1.14[W,Y,O] (-9%) | 22 (0%) | 24[Y,O] (4%) | 337[W,Y,O] (-42%) | 289[W,Y,O] (-40%) | 1.45[W] (-6%) | 1.38[W] (-16%) | 16[W] (-6%) | 16[W] (-11%) |
| avg | 150 | 109 | 1.36 | 1.28 | 22 | 25 | 451 | 396 | 1.46 | 1.42 | 16 | 16 |

Notes: wd = weekday; we = weekend. Superscripts indicate statistically significant differences between the restriction level in each row and the other three levels (p = .05).

In general, for both services, more trips are performed during weekdays than on weekends irrespective of the COVID-19 restrictions. This might suggest that both services are more likely to be used for utilitarian trips than recreational trips.

Concerning the number of daily trips, both services show a decreasing trend as the restrictions level becomes more severe. However, while dockless bike sharing usage dramatically drops under yellow, orange and red restriction levels (about -60% on weekdays and -70% on weekends), dockless e-scooters usage shows a smoother reduction, that becomes important only when restrictions are the tightest (-42% on weekdays, and -40% on weekends). Statistical tests support these findings: the average number of daily trips performed with e-scooters is significantly lower during red limitations, whereas it is not significant among other restriction levels. On the contrary, the reduction in bike sharing average daily trips is statistically significant between the white level and all the other more severe restriction levels. This seems to contradict the findings from the previous section, where higher fluctuations were observed for e-scooters. However the two analyses are looking at data from different perspectives: while the former analysis of time series separately considers each period characterized by constant restriction levels, here the temporal dimension is ignored and all such periods are jointly considered. Therefore, the present analysis is exclusively focusing on the effect of restriction levels and shows that these are relatively less impacting for shared e-scooters, while the opposite is true for other exogenous factors that were analyzed in the previous subsection. For the same reason, bike sharing daily trips during red restrictions are higher than in less restrictive periods, given their relatively high levels of use in November and the lack of a sharp increase of their use at the beginning of 2021.

Looking at the average Euclidean distances between trip ends, the figures show that shared e-scooters are used to cover longer distances than shared bikes. Differently from the average number of daily trips, such distances do not show a common trend. Dockless bike sharing are used to cover longer distances on weekdays than on weekends during the period with the least restriction level. However the variation under different restrictions changes according to the day type. On the one hand, on weekdays, as restrictions become stricter, the average trip distance reduces significantly during yellow (-6%) and red restrictions (-9%), but not during orange restrictions (-3%). On the other hand, weekends' average trip distance significantly increases moving from white towards yellow and orange restrictions (+9% and +11%, respectively), as well as it significantly drops during red restrictions (-9%). Apparently, in the period when red limitations were in force, dockless shared bikes were used more but for shorter trips than in yellow and orange restrictions.

Dockless shared e-scooters are used to cover longer distances on weekends than on weekdays when the level of restrictions is white, consistently with their more frequent use for non-systematic trips that was already noted. As the restrictions increase, the average trip distance reduces. More marked reductions can be observed during weekends (-16%, -13%, and -16% in yellow, orange and red limitations, respectively) than on weekdays, however in both cases the difference is significant compared to the least level of restrictions.

Regarding the average trip duration, dockless bike sharing trips last more than trips performed with shared e-scooters, albeit the latter are used to cover longer distances. It is indeed common experience that e-scooters travel faster than muscular bikes (and especially shared bikes whose ergonomic design seems not optimized to cycle fast). We also checked where the above mentioned filters applied to the maximum trip duration of the two services had an impact; however, when setting the maximum duration filter to 120 minutes for both services, we did not observe significant differences in our results.

We also note that the average durations estimated for the period with the least level of restrictions (W) are rather different from those reported in a published Italian report (*9*), where average durations for both services were quantified in 9.6 minutes; yet, data analyzed in such report refer to a different period (summer 2020) and include trips performed in several Italian's cities (not only in Turin), in which land use patterns might influence trip distances and hence duration.

It is also worth noting that both services are used for longer trips (in terms of duration) during weekends, even if origin and destination points are slightly closer in most cases. This might suggest a more relaxed attitude of drivers of both services during weekends, that could even choose longer and presumably safer paths in such circumstance.

Observing the usage of dockless bike sharing under different levels of limitations, the average trip duration does not significantly change during weekdays: such finding, combined with shorter distances between origin and destination, might suggest the choice of longer paths. On the contrary, the average trip duration significantly increases during weekends which is consistent with the increase on traveled distances as restrictions become tighter.

Finally, e-scooters average trip duration significantly decreases as restriction levels increases.

**Changes in the spatial distribution of trip ends**

**Figure 2** and **Figure 3** respectively show the spatial distribution of trip ends of bike sharing and of e-scooters within their service operational area. We refer the interested reader to Chicco and Diana (*40*) for a preliminary comparison of how such distributions change between the two services, since the focus of the present paper is rather to understand if there are differences in how such distributions change due to different restrictions.

Keeping in mind that blue spots in the chart represent densities below the average and warm colors densities above the average with increasing concentrations, **Figure 2** unveils that the use of dockless shared bikes tends to concentrate more in the central parts of the city and along the main corridors when Y-O-R restrictions are in place compared to days with less restrictions (W). In fact, yellow areas in the top left chart of the figure are smaller. Turning our attention to hotspots, that can be defined as places where the density of origin/destination trip ends is more than the average plus 5 times the standard deviation, three of them are visible and quite well defined when travel restrictions are minimal (white days): the strongest one is around the main railway station Porta Nuova (TORINO P. NUOVA label in the map), followed by decreasing order of importance by a second one around the main campus of Turin Polytechnic (POLITO) and a third one around another railway station, namely Porta Susa (TORINO P. SUSA).

When travel restrictions come into place, the Porta Nuova hotspot seems less neat especially in yellow days and a higher concentration of trips can be seen in the "San Salvario" neighborhood that is located next to the south-eastern edge of the Porta Nuova hotspot. Changes in the POLITO hotspot are less marked, except for a stronger concentration in the POLITO site during red days (bottom-right chart) whose stronger restrictions did not affect commuters compared to non-red days in the same period.

When considering e-scooters spatial distributions of trip ends (**Figure 3**), the hotspots near POLITO is now missing and another one appears one of the main square of the city center, piazza Castello (PIAZZA CASTELLO label in the map), which is located next to northern edge of the Porta Nuova railway station. This is a confirmation of the previously noted differences in trip purposes among the two services, since e-scooters are less used for work and study trips and more for shopping, recreational and non-systematic trips. We also do not observe the same trend of larger yellow areas when restrictions become stronger that we noted for shared bikes. However, such higher density spots become more scattered around the city, in proximity of the main mobility attractors and of the more densely populated neighborhoods that are normally served by public transport. Indeed, public transport operations did not substantially change according to restriction levels for the period under consideration: nevertheless, it seems that e-scooters are somehow substituting transit at the peaks of the pandemic, probably because they are considered safer. If the top left map of **Figure 3** shows the strong complementarity between public transport and e-scooters, which is possibly also due to the vehicle relocation policy of e-scooters operators, the other three maps show also substitution effects when public transport is not any more felt safe enough.

Such strong link between public transport and e-scooters is confirmed when considering hotspots, according to the previous definition. When restriction levels increase, new red areas appear and are overlapping with the east-west path of the only metro line presently operating in Torino. In orange days, a new hotspot was even observed around another railway station  located in the northern part of the city (TORINO DORA).

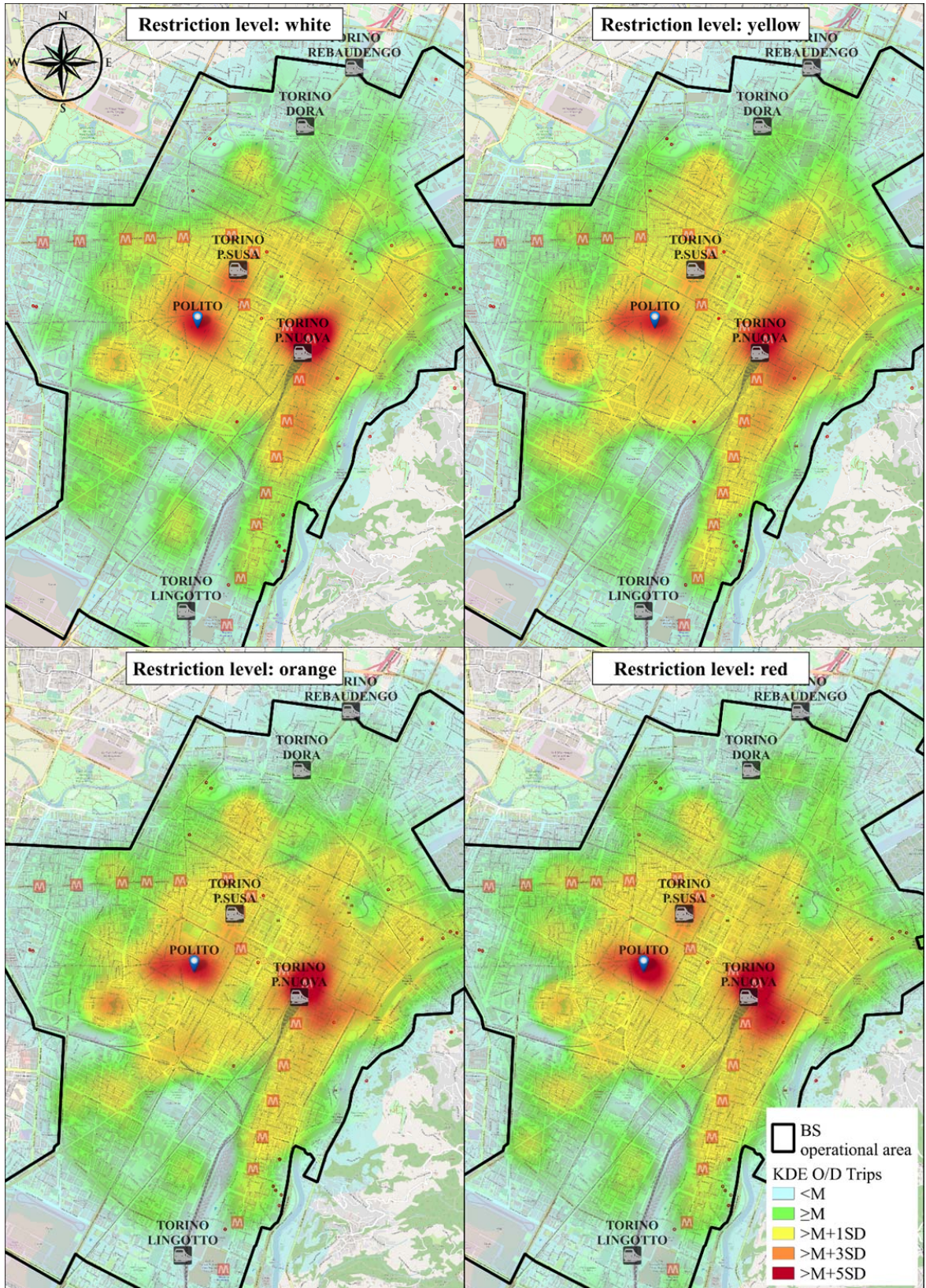

**Figure 2 Hotspot maps of dockless shared bikes for different restriction levels**

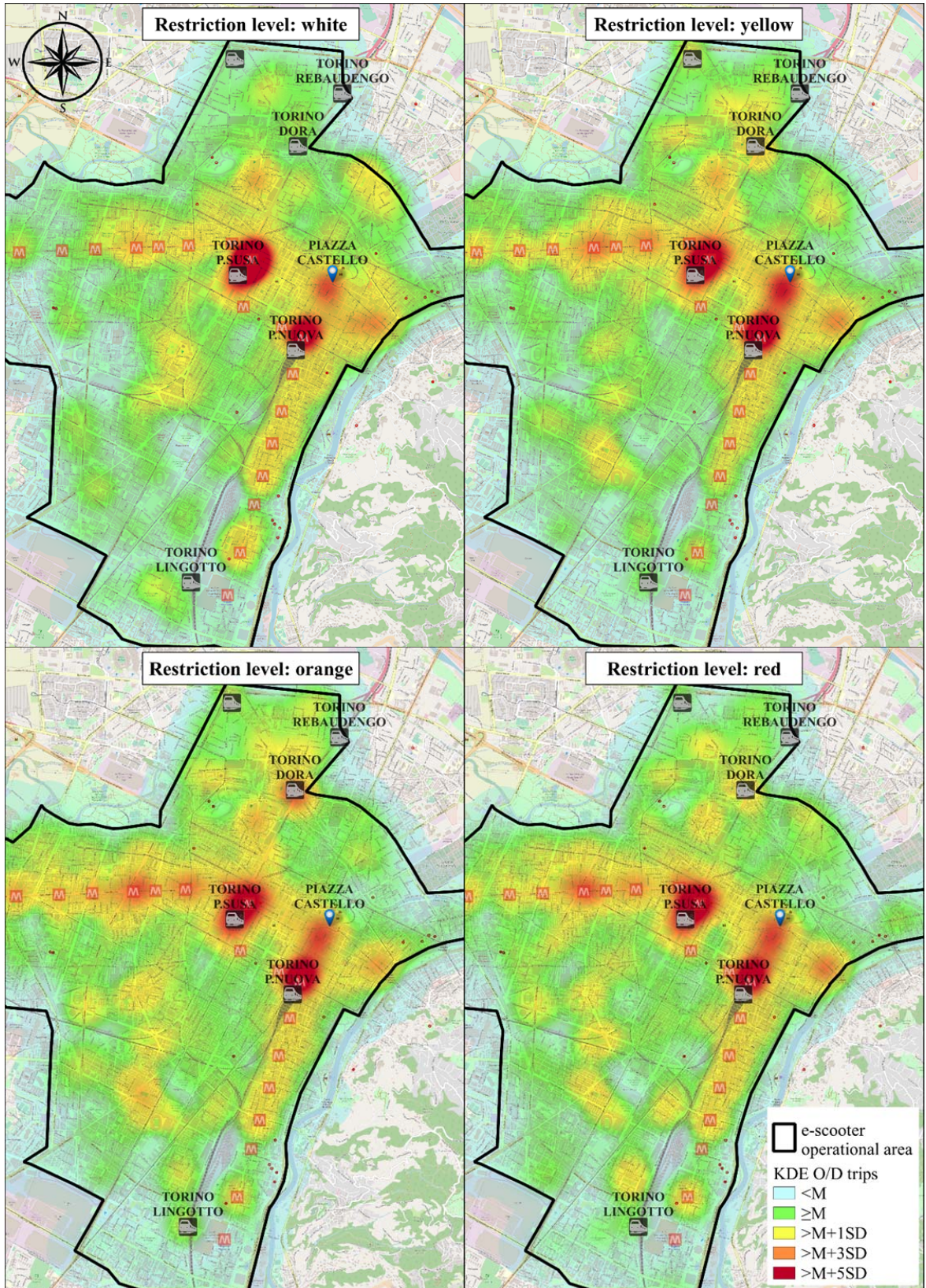

**Figure 3 Hotspot maps of e-scooters for different restriction levels**

**CONCLUSIONS**

Transportation systems all over the world have been severely hit by COVID-related travel restrictions. To avoid a massive modal diversion from public transport towards private car and to promote the adoption of more sustainable travel means, many cities worldwide have expedited the construction of cycle lanes (even temporary ones) and have strengthened the supply of shared micromobility services (*42*), in particular of e-scooters (*9*). Thus, despite the shutdown of several micromobility services around the world, shared micromobility might become an asset in many cities during the post-lockdown phase (*10*).

Yet, the analyses here presented have shown to which extent different shared micromobility services can be affected by various levels of restrictions, that could endanger their existence if such a shaky situation will continue in the future. Shared bikes and shared e-scooters patterns of use were observed in the city of Torino (Italy) under different travel restrictions between end 2020 and the beginning of 2021. The main findings are as follows:

- Shared bikes are more consistently used for systematic and work-related trips, thus in general suffering less from travel restriction levels more related to leisure trips.
- E-scooters experienced a larger variability in their demand, independently on travel restriction levels, and had a quicker recovery when travel restrictions were loosened.
- E-scooters seem more a complement or a substitute of urban public transport, according to the different restriction levels, where patterns of use of shared bikes are more independent.
- Beyond the decrease in the number of trips, also both average trip distances and durations decreased for e-scooters, which represents an additional loss in terms of revenues since shorter trips were clearly not compensated by an increased vehicle turnover. Concerning shared bikes, speeds were reduced during travel restriction periods, possibly related to a different travel attitude since travel restriction impacting more discretionary activities could have led to the opposite result.

Different extensions of the present work are sought to have a clearer picture of such patterns. Above all, it would be important to more clearly distinguish travel restriction from seasonal effects, ideally comparing pre-COVID-19 with COVID-19 data. Indeed many services, including most of Italian ones concerning e-scooters, were launched immediately before the spread of the disease so that "baseline" data are not available.

**AUTHOR CONTRIBUTIONS**

The authors confirm contribution to the paper as follows: study conception and design: Andrea Chicco and Marco Diana; data collection: Andrea Chicco. Author; analysis and interpretation of results: Andrea Chicco and Marco Diana; draft manuscript preparation: Marco Diana and Andrea Chicco. All authors reviewed the results and approved the final version of the manuscript.

## REFERENCES

1.  Scorrano, M., and R. Danielis. Active Mobility in an Italian City: Mode Choice Determinants and Attitudes before and during the Covid-19 Emergency. *Research in Transportation Economics*, Vol. 86, No. January, 2021. https://doi.org/10.1016/j.retrec.2021.101031.
2.  Schulte-Fischedick, M. The Implication of COVID-19 Lockdowns on Surface Passenger Mobility and Related CO2 Emission Changes in Europe. *Applied Energy*, Vol. 300, No. June, 2021, p. 117396. https://doi.org/10.1016/j.apenergy.2021.117396.
3.  Falchetta, G., and M. Noussan. The Impact of COVID-19 on Transport Demand, Modal Choices, and Sectoral Energy Consumption in Europe. *IaeeOrg*, 2020. https://doi.org/https://doi.org/10.1016/S1473-3099(20)30120-1.
4.  Eisenmann, C., C. Nobis, V. Kolarova, B. Lenz, and C. Winkler. Transport Mode Use during the COVID-19 Lockdown Period in Germany: The Car Became More Important, Public Transport Lost Ground. *Transport Policy*, Vol. 103, No. January, 2021, pp. 60–67. https://doi.org/10.1016/j.tranpol.2021.01.012.
5.  Aloi, A., B. Alonso, J. Benavente, R. Cordera, E. Echániz, F. González, C. Ladisa, R. Lezama-Romanelli, Á. López-Parra, V. Mazzei, L. Perrucci, D. Prieto-Quintana, A. Rodríguez, and R. Sañudo. Effects of the COVID-19 Lockdown on Urban Mobility: Empirical Evidence from the City of Santander (Spain). *Sustainability*, Vol. 12, 2020, p. 3870. https://doi.org/doi:10.3390/su12093870.
6.  NACTO. *Shared Micromobility in the U.S.: 2019*. National Association of City Transportation Officials, New York, 2020.
7.  U.S. Department of Transport. Effects of COVID-19 on Bikeshare (Docked and Dockless) and E-Scooter Operations. https://data.bts.gov/stories/s/Docked-and-Dockless-and-E-Scooter-System-Changes-2/kar5-6dpn#effects-of-covid-19-on-bikeshare-and-e-scooter-operations-march-through-august-2020. Accessed Jul. 12, 2021.
8.  Bergantino, A. S., M. Intini, and L. Tangari. Influencing Factors for Potential Bike-Sharing Users: An Empirical Analysis during the COVID-19 Pandemic. *Research in Transportation Economics*, Vol. 86, No. January, 2021, p. 101028. https://doi.org/10.1016/j.retrec.2020.101028.
9.  Ciuffini, M., S. Asperti, V. Gentili, R. Orsini, and L. Refrigeri. *4° Rapporto Nazionale Sulla Sharing Mobility 2019*. Osservatorio Nazionale Sharing Mobility, Roma. http://osservatoriosharingmobility.it/wp-content/uploads/2020/12/IV-RAPPORTO-SHARING-MOBILITY.pdf, 2020.
10. Dias, G., E. Arsenio, and P. Ribeiro. The Role of Shared E-Scooter Systems in Urban Sustainability and Resilience during the Covid-19 Mobility Restrictions. *Sustainability*, Vol. 13, No. 13, 2021, p. 7084. https://doi.org/10.3390/su13137084.
11. Wang, H., and R. B. Noland. Bikeshare and Subway Ridership Changes during the COVID-19 Pandemic in New York City. *Transport Policy*, Vol. 106, No. October 2020, 2021, pp. 262–270. https://doi.org/10.1016/j.tranpol.2021.04.004.
12. ITF. Safe Micromobility. *International Transport Forum Policy Papers*, No. 85, 2020. https://doi.org/10.1787/0b98fac1-en.
13. El-Assi, W., M. Salah Mahmoud, and K. Nurul Habib. Effects of Built Environment and Weather on Bike Sharing Demand: A Station Level Analysis of Commercial Bike Sharing in Toronto. *Transportation*, Vol. 44, No. 3, 2017, pp. 589–613. https://doi.org/10.1007/s11116-015-9669-z.
14. Trivedi, T. K., C. Liu, A. L. M. Antonio, N. Wheaton, V. Kreger, A. Yap, D. Schriger, and J. G. Elmore. Injuries Associated With Standing Electric Scooter Use. *JAMA network open*, Vol. 2, No. 1, 2019, p. e187381. https://doi.org/10.1001/jamanetworkopen.2018.7381.
15. Reiss, S., and K. Bogenberger. GPS-Data Analysis of Munich's Free-Floating Bike Sharing System and Application of an Operator-Based Relocation Strategy. *IEEE Conference on Intelligent Transportation Systems, Proceedings, ITSC*, Vol. 2015-Octob, No. 4, 2015, pp. 584–589. https://doi.org/10.1109/ITSC.2015.102.
16. Ramboll. *Achieving Sustainable Micro-Mobility*. 2020.

17. de Bortoli, A., and Z. Christoforou. Consequential LCA for Territorial and Multimodal Transportation Policies: Method and Application to the Free-Floating e-Scooter Disruption in Paris. *Journal of Cleaner Production*, Vol. 273, 2020, p. 122898. https://doi.org/10.1016/j.jclepro.2020.122898.

18. Chen, Z., D. van Lierop, and D. Ettema. Dockless Bike-Sharing Systems: What Are the Implications? *Transport Reviews*, Vol. 40, No. 3, 2020, pp. 333–353. https://doi.org/10.1080/01441647.2019.1710306.

19. PBOT. *2018 E-Scooter Findings Report*. Portland Bureau of Transportation, Portland, OR, 2018.

20. Reck, D. J., H. Haitao, S. Guidon, and K. W. Axhausen. Explaining Shared Micromobility Usage, Competition and Mode Choice by Modelling Empirical Data from Zurich, Switzerland. *Transportation Research Part C: Emerging Technologies*, Vol. 124, No. November 2020, 2021, p. 102947. https://doi.org/10.1016/j.trc.2020.102947.

21. Lee, M., J. Y. J. Chow, G. Yoon, and B. Y. He. Forecasting E-Scooter Substitution of Direct and Access Trips by Mode and Distance. *Transportation Research Part D*, Vol. 96, No. May, 2021, pp. 1–21. https://doi.org/10.1016/j.trd.2021.102892.

22. Laa, B., and U. Leth. Survey of E-Scooter Users in Vienna: Who They Are and How They Ride. *Journal of Transport Geography*, Vol. 89, No. October, 2020, p. 102874. https://doi.org/10.1016/j.jtrangeo.2020.102874.

23. Wang, K., X. Qian, G. Circella, Y. Lee, J. Malik, and D. Taylor Fitch. What Mobility Modes Do Shared E-Scooters Displace? A Review of Recent Research Findings. Presented at 100th Annual Meeting of the Transportation Research Board, Washington, D.C., , 2021.

24. Bieliński, T., and A. Ważna. Electric Scooter Sharing and Bike Sharing User Behaviour and Characteristics. *Sustainability (Switzerland)*, Vol. 12, No. 22, 2020, pp. 1–13. https://doi.org/10.3390/su12229640.

25. McKenzie, G. Spatiotemporal Comparative Analysis of Scooter-Share and Bike-Share Usage Patterns in Washington, D.C. *Journal of Transport Geography*, Vol. 78, 2019, pp. 19–28. https://doi.org/10.1016/j.jtrangeo.2019.05.007.

26. Chang, X., J. Wu, Z. He, D. Li, H. Sun, and W. Wang. Understanding User's Travel Behavior and City Region Functions from Station-Free Shared Bike Usage Data. *Transportation Research Part F: Traffic Psychology and Behaviour*, Vol. 72, 2020, pp. 81–95. https://doi.org/10.1016/j.trf.2020.03.018.

27. Bai, S., and J. Jiao. Dockless E-Scooter Usage Patterns and Urban Built Environments: A Comparison Study of Austin, TX, and Minneapolis, MN. *Travel Behaviour and Society*, Vol. 20, No. April, 2020, pp. 264–272. https://doi.org/10.1016/j.tbs.2020.04.005.

28. Caspi, O., M. J. Smart, and R. B. Noland. Spatial Associations of Dockless Shared E-Scooter Usage. *Transportation Research Part D: Transport and Environment*, Vol. 86, No. July, 2020, p. 102396. https://doi.org/10.1016/j.trd.2020.102396.

29. Khatri, R., C. R. Cherry, S. S. Nambisan, and L. D. Han. Modeling Route Choice of Utilitarian Bikeshare Users with GPS Data. *Transportation Research Record: Journal of the Transportation Research Board*, Vol. 2587, No. 1, 2016, pp. 141–149. https://doi.org/10.3141/2587-17.

30. Zou, Z., H. Younes, S. Erdoğan, and J. Wu. Exploratory Analysis of Real-Time E-Scooter Trip Data in Washington, D.C. *Transportation Research Record: Journal of the Transportation Research Board*, Vol. 2674, No. 8, 2020, pp. 285–299. https://doi.org/10.1177/0361198120919760.

31. Chen, E., and Z. Ye. Identifying the Nonlinear Relationship between Free-Floating Bike Sharing Usage and Built Environment. *Journal of Cleaner Production*, Vol. 280, 2021, p. 124281. https://doi.org/10.1016/j.jclepro.2020.124281.

32. McKenzie, G. Docked vs. Dockless Bike-Sharing: Contrasting Spatiotemporal Patterns. No. 114, S. Winter, A. Griffin, and M. Sester, eds., 2018, pp. 1–7.

33. Ciuffini, M., V. Gentili, D. Milioni, L. Refrigeri, G. Rossi, L. Soprano, and F. Squitieri. *2° Rapporto Nazionale Sulla Sharing Mobility*. Osservatorio Nazionale Sharing Mobility, Roma.

http://osservatoriosharingmobility.it/wp-content/uploads/2018/09/II-Rapporto-Nazionale_capitolo-dati_DEF_CON-INTESTAZIONE_5.pdf, 2018.

34. Li, A., P. Zhao, H. Haitao, A. Mansourian, and K. W. Axhausen. How Did Micro-Mobility Change in Response to COVID-19 Pandemic? A Case Study Based on Spatial-Temporal-Semantic Analytics. *Arbeitsberichte Verkehrs- und Raumplanung*, Vol. 1601, 2021. https://doi.org/10.3929/ethz-b-000473263.

35. Bontempi, E. The Europe Second Wave of COVID-19 Infection and the Italy "Strange" Situation. *Environmental Research*, Vol. 193, No. 110476, 2021. https://doi.org/10.1016/j.envres.2020.110476.

36. Mathew, J. K., M. Liu, S. Seeder, H. Li, and D. M. Bullock. Analysis of E-Scooter Trips and Their Temporal Usage Patterns. *ITE Journal*, Vol. 89, No. 6, 2019, pp. 44–49.

37. Kruskal-Wallis Test. In *The Concise Encyclopedia of Statistics*, Springer New York, New York, NY, pp. 288–290.

38. Silverman, B. W. *Density Estimation for Statistics and Data Analysis*. Chapman and Hall, London, 1986.

39. QGIS Development Team. QGIS 3.4.13. http://qgis.osgeo.org.

40. Chicco, A., and M. Diana. Understanding Micro-Mobility Usage Patterns: A Preliminary Comparison between Dockless Bike Sharing and e-Scooters in the City of Turin (Italy). *Transportation Research Procedia*, To appear, 2021.

41. Noland, R. B. Scootin' in the Rain: Does Weather Affect Micromobility? *Transportation Research Part A: Policy and Practice*, Vol. 149, No. February, 2021, pp. 114–123. https://doi.org/10.1016/j.tra.2021.05.003.

42. Lozzi, G., M. Rodrigues, E. Marcucci, T. Teoh, V. Gatta, and V. Pacelli. *Research for TRAN Committee - COVID-19 and Urban Mobility: Impacts and Perspectives*. european Parliament, Policy Department for Structural and Cohesion Policies, Brussels, 2020.